\documentclass[10pt,conference]{IEEEtran}
\IEEEoverridecommandlockouts
\usepackage{cite}
\usepackage{amsmath,amssymb,amsfonts}
\usepackage{algorithmic}
\usepackage{graphicx}
\usepackage{textcomp}
\usepackage{xcolor}
\usepackage{amsmath,amsfonts}
\usepackage{multirow}
\usepackage{algorithmic}
\usepackage{graphicx}
\usepackage{textcomp}
\usepackage{xcolor}
\usepackage{framed}
\usepackage{listings}
\usepackage{bbding} 
\usepackage{color}
\usepackage{enumitem}
\usepackage[many]{tcolorbox}
\usepackage{tcolorbox}
\usepackage{threeparttable}
\usepackage{colortbl}
\usepackage{pifont}
\usepackage{tikz}
\usepackage{booktabs}
\usepackage[font=small,skip=2pt]{caption}
\usepackage{hyperref}
\usepackage{cleveref}
\usepackage{url}
\usepackage{balance}

\usepackage{setspace}

\usepackage[font=small,skip=2pt]{caption}
\newcommand{\distance}{2pt}
\def\BibTeX{{\rm B\kern-.05em{\sc i\kern-.025em b}\kern-.08em
    T\kern-.1667em\lower.7ex\hbox{E}\kern-.125emX}}
\begin{document}

\title{Catch the Butterfly: Peeking into the Terms and Conflicts among
SPDX Licenses}

\author{
\IEEEauthorblockN{Tao Liu\IEEEauthorrefmark{1},
Chengwei Liu\IEEEauthorrefmark{2}\IEEEauthorrefmark{4},
Tianwei Liu\IEEEauthorrefmark{1},
He Wang\IEEEauthorrefmark{1},
Gaofei Wu\IEEEauthorrefmark{1},
Yang Liu\IEEEauthorrefmark{2},
Yuqing Zhang\IEEEauthorrefmark{1}\IEEEauthorrefmark{3}\IEEEauthorrefmark{4}
}
\text{liut@nipc.org.cn, 
chengwei.liu@ntu.edu.sg, zhangyq@nipc.org.cn}
\IEEEauthorblockA{\IEEEauthorrefmark{1} School of Cyber Engineering, Xidian University, Xian, China}
\thanks{\IEEEauthorrefmark{4} Corresponding authors.}
\IEEEauthorblockA{\IEEEauthorrefmark{2} School of Computer Science and Engineering, Nanyang Technological University, Singapore}
\IEEEauthorblockA{\IEEEauthorrefmark{3} University of Chinese Academy of Sciences, Beijing, China}

}

\maketitle

\begin{abstract}
The widespread adoption of third-party libraries (TPLs) in software development has significantly accelerated the creation of modern software. However, this convenience comes with potential legal risks. Developers may inadvertently violate the licenses of TPLs, leading to legal issues. While existing studies have explored software licenses and potential incompatibilities, these studies often focus on a limited set of licenses or rely on low-quality license data, which may affect their conclusions. To address this gap, there is an urgent need for a high-quality license dataset that encompasses a broad range of mainstream licenses and provides accurate terms and conflict information, to help developers navigate the complex landscape of software licenses, avoid potential legal pitfalls, and guide more informed and effective solutions for managing license compliance and compatibility in software development.

To this end, we conduct the first work to understand the mainstream software licenses based on term granularity and obtain a high-quality dataset of 453 SPDX licenses with well-labeled terms and conflicts. Specifically, we first conduct a differential analysis of the mainstream platforms that provide license data to understand the terms and attitudes of each license. Next, we further propose a standardized set of license terms to capture and label existing mainstream licenses with high quality. Moreover, we improve the existing license conflict mode to include copyleft conflicts and conclude the three major types of license conflicts among the 453 SPDX licenses. Based on the dataset, we carry out two empirical studies to reveal the concerns and threats from the perspectives of both licensors and licensees. One study provides an in-depth analysis of the similarities, differences, and conflicts among SPDX licenses, and the other revisits the usage and conflicts of licenses in the NPM ecosystem and draws conclusions that differ from previous work. Our studies reveal some insightful findings and disclose relevant analytical data, which set the stage for further research into the complexities of license compliance and compatibility.
\end{abstract}

\begin{IEEEkeywords}
open source, third-party library, license, compatibility, conflict, dependency
\end{IEEEkeywords}

\section{Introduction}
With the rapid growth of functional complexity in software applications, developers are increasingly incorporating third-party libraries (TPLs) as open-source dependencies. This approach avoids reinventing the wheel and facilitates rapid software development. According to \textit{Libraries.io}~\cite{librariesio}, over 6 million open source packages are published on 32 different package managers, and these TPLs are prevalent, manifesting in the majority (79\%) of current software applications as revealed by \textit{Contrast Security}~\cite{contrastsecurity}. 

However, every coin has two sides. While reusing TPLs reduces development efforts, it has also placed potential legal risks of violating license compatibility due to the lack of legal expertise. Specifically, TPLs are usually published with licenses. For instance, NPM assigns \textit{ISC} license by default when publishing packages~\cite{makari2022prevalence}. Software licenses usually state the rights granted to the licensee, as well as the limitations to be observed and the obligations to be fulfilled in the exercise of those rights. Legal issues arise when these requirements are not met. Moreover, such threats can propagate through dependency relations, resulting in greater impact and causing economic loss~\cite{schoettle2019open, Openisnotequaltopublic, 2004software}. For instance, Cisco has been sued by the FSF~\cite{FreeSoftwareFoundation} for violating the licenses of numerous programs under FSF's copyright purview, such as GCC, binutils, and the GNU C Library. These programs are licensed under the GPL license~\cite{gpl}, mandating that derivative works adhere to the principles of open source. Cisco's non-compliance with this mandate led to substantial economic losses~\cite{cisco}.

Such cases have warned developers and users of the potential legal threats when reusing third-party components. 
Many researchers \cite{choosealicense, tldr, diaochan, azhakesan2020sharing, kapitsaki2015open, kapitsaki2017automating, pfeiffer2022license, wang2021novel, ShiQiu2021empirical, xu2023liresolver, xu2021lidetector} have studied the use and compatibility of licenses from different aspects. However, most of these works~\cite{azhakesan2020sharing, kapitsaki2015open, kapitsaki2017automating, wang2021novel, ShiQiu2021empirical} are conducted at license level, based on limited licenses and conflict relations that are known to the public, which could largely compromise the insights derived from their studies. To unveil the compatibility issues hidden in
customized licenses, Xu et al.~\cite{xu2023liresolver, xu2021lidetector} investigated the term-level license incompatibility by training NLP models to retrieve license terms, which unveils much more potential incompatibilities compared to other works. However, they focus primarily on customized licenses, and the training data is directly from existing license platforms, which is confirmed to be unreliable according to our analysis in \Cref{subsec:licenselabeling}. 
Moreover, many software composition analysis (SCA) tools \cite{imtiaz2021comparative} also started to claim support for license-related detection and analysis. However, these SCA tools only involve the coarse-grained identification of license information (license name and
quantity) and do not support fine-grained license analysis (term extraction), not to mention license compatibility detection. 
Therefore, a well-labeled dataset of terms for mainstream software licenses could definitely benefit the understanding of the communities and promote actionable solutions to avoid legal issues related to license incompatibility.

However, it is non-trivial to construct such a high-quality dataset for mainstream licenses, we mainly face the following challenges: 
\textbf{1) Lack of standardized license model.} There lack of consensus and standards on the license terms. For a given license text, different people could retrieve the content into different sets of terms. For instance, Choosealicense~\cite{choosealicense}, TLDRLegal~\cite{tldr}, and OpenEuler~\cite{diaochan} have different item sets with 16, 23, and 18 license terms.
\textbf{2) The ambiguity of license text.} Software licenses are often composed of intricate declarations, which encompass numerous legal provisions and are difficult for non-professionals to comprehend.
\textbf{3) Lack of license conflict model.} Although existing works have collected some pairs of license conflicts from known sources, there still lacks an in-depth understanding of the major types of license conflicts, to guide the construct of license conflict dataset for mainstream licenses. 
\textbf{4) Irregular format when assigning licenses.} We notice a casual practice that many developers use irregular license identifiers when assigning licenses to artifacts, which makes it difficult to accurately identify all licenses of all TPLs. 

To overcome these challenges, 1) we first study the term definitions in three mainstream platforms (Choosealicense~\cite{choosealicense}, TLDRLegal~\cite{tldr} and OpenEuler~\cite{diaochan}) for license analysis. Based on the differential analysis and manual checking with lawyers, we derive a set of 22 license terms as the fundamental structure of defining a software license.
2) We further label the mainstream SPDX licenses based on our term set, study the major inconsistencies of these three platforms, and construct a term dataset for SPDX licenses, which is also well validated with high quality (83.68\% consistency among 9 manual examiners). 
3) Moreover, we improve the existing incompatibility modes~\cite{xu2021lidetector} by adding copyleft license conflicts that are decided by the infectious copyleft nature. Based on this, we investigate the possible license conflicts by pairwise comparison of their license terms, and construct a dataset for three major types of license conflicts. 
4) Based on these, we further conduct two empirical studies. Specifically, the first study interprets the similarities and major differences among mainstream licenses, which unveils the main concerns and requirements of licensors. Moreover, it also interprets the differences in licenses from the perspective of potential conflicts. The second study stands on the aspects of licensees, as an ecosystem recognized as relatively safe on license compatibility by existing work~\cite{ShiQiu2021empirical}, we revisit the license usage, license changes, and the potential threats of license conflict in the NPM ecosystem based on our datasets of license terms and conflicts, for a more in-depth understanding of license conflicts hidden in TPL ecosystems.

Our comprehensive research uncovers significant gaps among existing sources of license labeling. These platforms adopt different sets of license terms when interpreting software licenses. Most SPDX licenses are mislabeled by these platforms, averagely each license contains 10.3 terms while 2.7 of them are mislabeled. This discrepancy underscores the complexity and ambiguity surrounding software license interpretation. From the perspective of licensors, there exists a clear consensus on the rights and obligations when designing licenses. Rights like \textit{Sublicense} and some obligations usually as preconditions to \textit{Distribute} bring the major diversity of licenses, which also contribute to the most terms-based license conflicts. \textit{Use Patent Claim} is usually granted in copyleft licenses to avoid legal disputes, and interestingly, it brings the most copyleft license conflicts. Despite the NPM ecosystem is dominated by the \textit{MIT}, \textit{ISC} and \textit{Apache-2.0} licenses, there are still considerable copyleft licenses and other minority licenses. Some developers choose to switch their licenses to these minority licenses instead of \textit{MIT} and \textit{ISC}. However, these minority licenses exacerbate conflicts with the most common licenses. For instance, 67.8\% of right conflicts are related to \textit{MIT} license as it grants \textit{Sublicense} to licensees, and 85.3\% of obligation conflicts are caused by the \textit{Inlcude Notice} and \textit{State Changes} in \textit{Apache-2.0}.  

In summary, we make the main contributions as follows.

\begin{itemize}[leftmargin=*]
\item We conducted the first large-scale analysis to clarify the granted permissions and required obligations by differential analysis. Based on this, we proposed a standardized term set to capture the mainstream licenses and a high-quality dataset containing 453 SPDX licenses with well-labeled attitudes on each license term.
\item We improved the existing license conflict modes to include infectious copyleft conflicts. Based on our dataset, we also constructed a detailed mapping of three types of license conflicts between any two SPDX licenses.
\item We carried out two empirical studies to unveil the concerns and potential threats from both perspectives of licensors and licensees, including \ding{172} an empirical study on the similarities, differences, and conflicts among SPDX licenses, and \ding{173} an in-depth revisit to the license usage and conflicts in the NPM ecosystem. Our findings can be utilized to mitigate potential license conflicts for developers, as well as shed light on future research.
\item We have open-sourced the datasets on our website~\cite{ourwebsite} to facilitate relevant research on the licenses.
\end{itemize}

\section{Background and Motivating example}
\label{section2}

\subsection{Background}
\label{subsec:background}

\begin{figure}[]
\center
\includegraphics[width=0.9\columnwidth]{"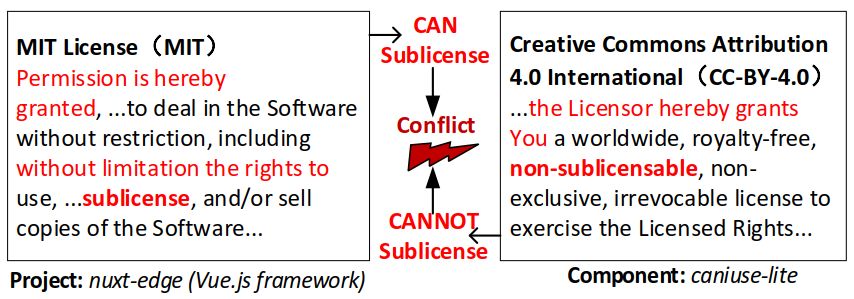"}
\caption{Examples of conflicts between licenses.}
\label{fig:conflict case}
\end{figure}

\textbf{Software licenses} are usually described in the form of free text in user projects and distributed along with the software, in which the owner grants the licensees what it is permitted to do (rights) and what it must comply with (obligations)~\cite{softwarelicenseUrl}. Specifically, rights and obligations are usually described as \textbf{license terms}~\cite{licensetermUrl} to state different behaviors related to licensing. For each specific license term, different licenses could have different \textbf{attitudes}~\cite{xu2021lidetector}, i.e., permission (CAN) and prohibition (CANNOT) of rights, or compulsory (MUST) on obligations. When software artifacts with different licenses are used together, the different attitudes of the different licenses towards the same term may lead to conflicts.

\subsection{Motivating Example}
Here we use an example of \textit{nuxt-edge}, a popular and minimalistic framework for server-rendered Vue.js applications, to illustrate how license conflicts are introduced. As depicted in \textbf{Fig.} \ref{fig:conflict case}, \textit{nuxt-edge} is released under \textit{MIT license}, while it has a third-party dependency \textit{caniuse-lite} is released under \texttt{CC-BY-4.0} license. According to the license text, \textit{MIT license} permits licensees \textit{Sublicensing} to third-party while this is not allowed in \texttt{CC-BY-4.0}. In this case, the conflict between licenses occurs and can have a large impact due to the popularity of \textit{nuxt-edge}.

License compatibility issues are not new to the open-source communities, many existing Software Composition Analysis (SCA) tools have begun to claim support for detecting licenses and potential conflicts of user projects. In this context, we further conducted a case study on the ability of existing SCA tools. Specifically, we collect 8 well-known SCA tools, 4 of which (FOSSA \cite{FOSSA}, Fossology \cite{FOSSology}, Askalono \cite{Askalono} and ScanCode \cite{ScanCode}) are freely available and support license detection according to their documents, and 4 of which we are unable to find their free trials. We detect the licenses and compatibility issues in \textit{nuxt-edge} by these tools. \Cref{tab:newSCA} presents the test results. 

It can be seen that neither Fossology nor Askalono detects the dependency information for \textit{nuxt-edge}, while
FOSSA and ScanCode not only list the dependency information,
but also includes package \textit{caniuse-lite} in the above conflict example. In addition, all four tools detected the license \textit{MIT} of package \textit{nuxt-edge}, in particular, FOSSA accurately detected the license \textit{CC-BY-4.0} of dependency package \textit{caniuse-lite}. Therefore, it can be
seen that existing SCA tools only involve the coarse-grained
identification of license information (license name and quantity),
but not the fine-grained analysis of license terms (term extraction
and compatibility judgment). Although the FOSSA tool detected 8
terms, they are all without attitude and do not correspond to a specific license, so it is impossible to judge the compatibility through
term analysis, but only give reminders on the use of some dependency packages.

\begin{table}[]
\centering
\caption{Comparison of SCA tool tests for conflict example}
\label{tab:newSCA}
\scalebox{0.65}{
\begin{tabular}{cll}
\hline
SCA Tools & \multicolumn{1}{c}{Dependency}                                                                                                            & \multicolumn{1}{c}{License}                                                                     \\ \hline
FOSS      & \begin{tabular}[c]{@{}l@{}}Number: 862\\ Name: caniuse-lite and others\\ License: CC-BY-4.0 and others\\ Issues: 8 reminders\end{tabular} & \begin{tabular}[c]{@{}l@{}}Number: 19\\ Name: MIT and others\\ Term: 8 obligations\end{tabular} \\ \hline
Fossology & No results                                                                                                                                & \begin{tabular}[c]{@{}l@{}}Number: 7\\ Name: MIT and others\end{tabular}                        \\ \hline
Askalono  & No results                                                                                                                                & \begin{tabular}[c]{@{}l@{}}Number: 2\\ Name: MIT\end{tabular}                                   \\ \hline
ScanCode  & \begin{tabular}[c]{@{}l@{}}Number: 1355\\ Name: caniuse-lite and others\end{tabular}                                                      & \begin{tabular}[c]{@{}l@{}}Number: 16\\ Name: MIT\\ Copyright Info: 6\end{tabular}              \\ \hline
\end{tabular}}
\end{table}

We know SCA tools work on pre-existing knowledge bases, and FOSSA has its own license dataset (i.e., TLDRLegal~\cite{tldr}) that provides license terms and attitudes for popular licenses, we further investigate the data quality of TLDRLegal. Specifically, we find another two mainstream platforms providing similar license labelings, and compare their terms. We find significant differences among these platforms, these platforms not only have different license term sets but also place different attitudes when labeling license terms (more details are discussed in ~\Cref{sec:Approach}). Therefore, it is vital to have a well-labeled license dataset with reasonable terms and precise attitudes before precisely analyzing license compatibility.

\section{License Term Analysis}
\label{sec:Approach}

\begin{figure}[]
\center
\includegraphics[width=0.4\textwidth]{"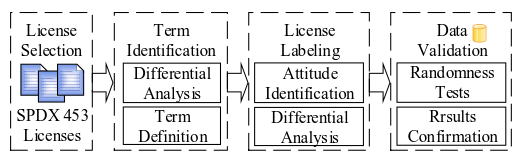"}
\caption{The overview of data processing on term extraction.}
\label{fig:overviewofdata}
\end{figure}

In this section, we present the detailed steps for collecting license terms and attitudes, as well as the measures to assure the accuracy of data acquisition. \textbf{Fig.} \ref{fig:overviewofdata} presents the overview of data processing on term extraction, including license selection, term identification, attitude labeling, and data validation.

\subsection{License Selection}
We first determine the scope of licenses we aim to investigate. 
Some non-profit organizations have proposed different recommended license lists to guide the community in selecting licenses. For instance, the Free Software Foundation (FSF)~\cite{FreeSoftwareFoundation} has concluded a list of 161 licenses for free software, and the Open Source Initiative (OSI)~\cite{osiUrl} has also released a list of 114 licenses that comply with the open source principles. 
However, these lists are specifically recommended for particular developers, instead of the general purpose of reusing existing third-party components in modern software development. In this case, we select the license list released by Software Package Data Exchange (SPDX) that is specifically designed as a standard to regulate the exchange of software bill materials as fundamental licenses to investigate. Until the time of our data collection, the license list of SPDX contains 453 different licenses, including not only all 173 licenses concluded by the FSF and OSI but also 280 additional commonly used licenses. To this end, we collect the metadata of the 453 SPDX licenses, including their standard name, abbreviations, and full content body.

\subsection{License Term Identification}
\label{subsec:term_identification}

Extending license detection from simply comparing licenses by names to term-based conflict detection has proven necessary and effective~\cite{xu2021lidetector,xu2023liresolver}. However, existing work fails to examine the accuracy of terms labeled by existing license data sources, leading to the inaccuracy of license conflict detection (as explained in~\Cref{subsec:licenselabeling}). Therefore, we extract license terms based on the full-text content of the license.

\begin{table*}[]
\centering
\caption{The inclusion of license terms among 3 mainstream platforms}
\footnotesize
\label{tbl:term_consistency}
\scalebox{0.7}{
\begin{tabular}{c|ll|ll|ll}
\toprule
\multicolumn{1}{l|}{}                   & \multicolumn{2}{c|}{\textbf{All Included}}                                & \multicolumn{2}{c|}{\textbf{Included in 2 platform}} & \multicolumn{2}{c}{\textbf{Unique Terms}}            \\ 
\midrule
                                       & \textbf{Distribute}                             & Private Use             & Use Patent Claims   &                                & Relicense              &                             \\
                                       & \textbf{Modify}                                 & \textbf{Use Trademark}  &                     &                                & Sublicense             &                             \\
                                       & \textbf{Commercial Use}                         & \textbf{Place Warranty} &                     &                                & Statically Link        &                             \\
\multirow{-4}{*}{\textbf{Rights}}      & \textbf{Hold Liable}                            &                         &                     &                                &                        &                             \\ 
\midrule
                                       & \textbf{Include Copyright}                      & Disclose Source         & Include Original    & Contact Author                 & Include Notice         & \underline{Pay Above Use Threshold}     \\
                                       &  \textbf{Include License} & \textbf{State Changes}  & Give Credit         & Include Install Instructions   & Rename                 & \underline{Same License}                \\
\multirow{-3}{*}{\textbf{Obligations}} &                                                 &                         &                     &                                & Compensate for Damages & \underline{Network Use Is Distribution} \\
\bottomrule
\end{tabular}
}
\begin{enumerate}
    \item Bold terms are inconsistent among platforms, and underlined terms are excluded from our final term set. For details of terms, please visit our website~\cite{ourwebsite}.
\end{enumerate}
\end{table*}

Before interpreting the license text, we first investigate three existing mainstream platforms (Choosealicense~\cite{choosealicense}, TLDRLegal~\cite{tldr}, and OpenEuler~\cite{diaochan}) that provide license term labelings, to obtain a general understanding of license terms. As presented in~\Cref{tbl:term_consistency}, a total of 25 different license terms (11 rights and 14 obligations) have been identified in these platforms. According to our observation, the inclusion of license terms varies on these platforms. Only 7 right terms and 4 obligation terms are included by all three platforms, and 9 terms are unique terms that are only defined by one platform. In addition, the interpretation of the same license terms also differs. In the 11 license terms included by all three platforms, the interpretations of 9 terms are inconsistent (bold terms in~\Cref{tbl:term_consistency}). For instance, for the right term \textit{Use Trademark}, Choosealicense only mentions the right to use trademarks, while both TLDRLegal and OpenEuler also include the right to use the contributor's name and logo. Moreover, we also find that the definition of the term \textit{Include License} even varies in all three platforms.

To this end, 3 co-authors revisit these license terms with examples of 453 SPDX licenses to determine the set of terms and their corresponding definitions. Specifically,\ding{172} for terms that are included in all three platforms and have consistent definitions (2 terms), we basically follow the existing definitions; \ding{173} for terms that are included in all three platforms but have inconsistent definitions (9 terms), we vote on which definition to follow; \ding{174} for terms that are only included by two platforms (5 terms), we review the descriptions and find that all of them are consistent, 4 terms are accepted and 1 term is revised by us due to improper definition; \ding{175} for terms that are only included by single platforms (9 terms), we review the necessity of keeping these terms, 3 obligation terms are excluded (underlined terms in \Cref{tbl:term_consistency}). The detailed terms for each platform and our final term set are available on our website~\cite{ourwebsite}. We also conclude the major factors that result in the inconsistency of terms as follows.

\noindent $\bullet$ \textbf{Differences in the scope of licensed objects (5 terms).} The most common factor is the scope of the licensed object. For example, the licensed objects of \textit{Distribute} and \textit{Modify} are defined as ``original or derivative work" by TLDRLegal and OpenEuler, while Choosealicense only mentions ``licensed material" and ignores the derivatives.

\noindent $\bullet$ \textbf{Terms are aggregated (4 terms).} We notice 2 pairs of terms are aggregated by OpenEuler, \ding{172} \textit{Hold Liable} \& \textit{Place Warranty}, \ding{173} \textit{Include Copyright} \& \textit{Include License}. However, these pairs do not co-exist in considerable licenses, as discussed in ~\Cref{subsec:conflictstudy}.

\noindent $\bullet$ \textbf{Multiple Definition (1 term)}. For \textit{Include Notice}, we identify 12 different descriptions from TLDRLegal, which indicate different notice statements.

\noindent $\bullet$ \textbf{Confusion of rights and obligations (1 term).} For \textit{Contact Author}, both TLDRLegal and OpenEuler interpret it as the ability to contact the author about usage, while it is more like an obligation to contact the author for permission.

Moreover, 3 terms are excluded. \ding{172} \textit{Pay Above Use Threshold} is used by TLDRLegal, but we do not find it in any SPDX licenses. \ding{173} \textit{Same License} is excluded because it is a unique feature of copyright licenses, we analyze it separately in~\Cref{sec:licenseconflict}. \ding{174} \textit{Network Use Is Distribution} can be also included by \textit{Distribute}. After reviewing our results with 2 lawyers from our industry partner, 22 license terms are finally selected as the standardized term set.

\begin{table*}
\centering
\caption{The inconsistent license term labeling of three mainstream platforms against ground truth}
\scalebox{0.6}{
\begin{tabular}{ll|rrr|rrr}
\toprule
\textbf{Reasons of Misslabeling}                                                & \textbf{Categories}                                & \textbf{Choossalicense} & \textbf{TLDRLegal} & \textbf{OpenEuler} & \multicolumn{3}{c}{\textbf{\#/\%}}                                   \\ \toprule
\multirow{4}{*}{\textbf{(1) Poor Interpretation of License Text}} &  Described but not identified          & 7              & 245         & 180       & \multicolumn{2}{c}{432 / 33.51\%}      & \multirow{4}{*}{675 / 52.37\%}     \\
                                                       & Not described but labeled                     & 6              & 108          & 112       & \multicolumn{2}{c}{226 / 17.53\%}      &                                  \\
                                                       & Opposite attitude    & 0              & 8           & 2        & \multicolumn{2}{c}{10 / 0.78\%}        &                                  \\
 & Right and obligation confused               & 0              & 4           & 3         & \multicolumn{2}{c}{7 / 0.54\%}      &                                      \\                                                    
 \midrule
\textbf{(2) Missing Term Definitions}                      & N.A.                                          & 69             & 0           & 336       & \multicolumn{2}{c}{-}  & 405 / 31.42\%                                         \\ 
\midrule
\multirow{3}{*}{\textbf{(3) Inconsistent Explanation of License Terms}}     & Different scope of terms    & 6              & 129         & 40        & 175 / 13.58\%          & \multicolumn{2}{c}{\multirow{3}{*}{197 / 15.28\%}} \\
                                                       & Multiple definitions       & 0              & 17          & 0         & 17 / 1.32\%            & \multicolumn{2}{c}{}                             \\
                                                       & Similar text to different terms & 0              & 5           & 0         & 5 / 0.38\%             & \multicolumn{2}{c}{}                             \\
                                                       \midrule

\textbf{(4) Ignoring License Terms from license names}                     & N.A.                                          & 0              & 1           & 11        & \multicolumn{2}{c}{-} & 12 / 0.93\%                                           \\
\midrule
\multicolumn{2}{c}{Total/Proportion}                                                                   & 88 / 6.83\%      & 517 / 40.11\% & 684 / 53.06\%  & \multicolumn{2}{c}{-} & 1,289 / 100\%                                          \\ \bottomrule
\end{tabular}}
\label{tab:reason_of_difflabels}
\end{table*}

\subsection{SPDX License Labeling}
\label{subsec:licenselabeling}

Based on the term set, we further label the attitude of each license term in the 453 SPDX licenses. For the right terms, we retrieve the permission as CAN and CANNOT, while for the obligation terms, we think there would be no negative attitude of obligations since they would be treated as forbidding specific rights in that case. We then retrieve the labeled license terms for each license from three platforms. For license terms that are labeled consistently, we take them as ground truth. However, we also notice that the labeling of most licenses varies significantly, and for them, 3 co-authors and 2 lawyers from our industry partner manually inspect the official license texts and determine the existence of each license's terms and their corresponding attitudes. The detailed ground truth of the license terms, as well as the labeling data from three platforms, are listed on our website~\cite{ourwebsite}.

Our results show that the quality of term labeling in these platforms is not only affected by the low coverage of SPDX licenses, but also by high inaccuracy compared with each other. \ding{172} Choosealicense, TLDRLegal, and OpenEuler only contain 41, 149, and 285 SPDX licenses, respectively. 165 SPDX licenses (36.42\%) are not indexed by any of these platforms. \ding{173} There also exists large portions of terms mislabeled on each platform (1,289 terms mislabeled in total). On average, 2.7 (26\%) license terms are mislabeled for each license by these platforms (each license contains 10.3 terms on average). Moreover, we inspect the inconsistency of license term labeling and summarize the reasons in~\Cref{tab:reason_of_difflabels}.

\noindent \textbf{(1) Poor interpretation of license text (675 mislabeling, 52.37\%).} 432 mislabeled terms (33.51\%) 
are explicitly described in the license text while not labeled. For instance, \textit{Sublicense} is explicitly mentioned by \texttt{APL-1.0} \cite{APL-1.0} but TLDRLegal does not label it. Furthermore, 226 mislabeled terms (17.53\%) are not mentioned in the license text but labeled, such as the \textit{Patent Use} of \textit{LGPL-3.0-only} on Choosealicense~\cite{lgpl-3.0}. Moreover, 10 mislabeled terms are labeled with the opposite attitudes, and 7 mislabeled terms are because of the confusion of rights and obligations. For example, for license text ``When you want to change the permissions of software, you must contact the author", both TLDRLegal and OpenEuler labeled it as CAN \textit{Contact Author}, while according to their definition (``Describes your ability to contact the author''), they twisted the original meaning and converted this obligation into a right.

\noindent \textbf{(2) Missing Term Definitions (405 mislabeling, 31.42\%).} As unveiled in~\Cref{subsec:term_identification}, these platforms use different term sets to interpret licenses. Therefore, if terms are not included by platforms, they can never be labeled. For instance, the \textit{Sublicense} in \texttt{APL-1.0} \cite{APL-1.0} are not labeled by Choosealicense and OpenEuler since they do not have such term definitions.

\noindent \textbf{(3) Inconsistent explanation of license terms (197 mislabeling, 15.28\%).} 
Typically, \ding{172} 175 mislabeled terms are due to the different scope of terms. For instance, the \textit{Use Trademark} is interpreted as the right to use trademarks by Choosealicense, while the other two platforms also include the contributor's name and logo. \ding{173} 17 mislabeled terms are introduced by multi-definition of terms (\textit{Include Notice}). \ding{174} For another 5 mislabeled terms, some platforms interpret similar texts into different license terms. For instance, \texttt{Fair}~\cite{fair} and \texttt{AAL}~\cite{aal} both contain similar text on warranty, but they are labeled as \textit{Place Warranty} and \textit{Hold Liable} by TLDRLegal, respectively.

\noindent {\textbf{(4) Ignoring License Terms from license names (12 mislabeling, 0.93\%).} Some licenses also endow some terms by license names, such as the CC-family licenses~\cite{Creativecommons}. For instance, \textit{Commercial Use} is not mentioned in the text of \texttt{CC-BY-NC-3.0}, but the \textit{NC} in its license name indicates "Non Commercial", which is neglected by OpenEuler.

\begin{tcolorbox}[size=title,opacityfill=0.2,breakable,boxsep=1mm]
\textbf{Remarks:} There exists significant gaps among existing license platforms on the understanding of software licenses, not only in the scope of license terms adopted but also in the interpretation of these terms for each license. \ding{172} Only 11 license terms are commonly adopted by three mainstream platforms, and only 2 of them are interpreted to the same definitions. \ding{173} Most SPDX licenses indexed by these platforms contain mislabeled terms. On average, each license contains 10.3 terms, while 2.7 of them are mislabeled terms. \ding{174} We redefine the license term with high confidence by aggregating opinions from existing license platforms, co-authors, and lawyers from our industry partner, as provided on our website~\cite{ourwebsite}. As a fundamental dataset, we believe more fine-grained license analysis can be further promoted.
\end{tcolorbox}

\subsection{Validation of terms and attitude labeling}
\label{3.4}

To validate the quality of term and attitude labeling, we randomly send out questionnaires to our colleagues, who are researchers on software engineering and open source security. Specifically, the questionnaires contain the terms set we defined and the full text of 10 randomly picked SPDX licenses, and we invite them to label the terms of each license and
their corresponding attitudes. As a result, we collected responses from 10 colleagues, one of which was discarded without following the requirements. 
These 90 licenses from 9 questionnaires consist of 864 terms in our dataset, the colleagues accurately labeled 723 terms (83.68\%), achieving high precision (84.1\%) and recall (83.7\%). 

After informing the participants and going through all their mislabeled terms, it was found that the mislabeled terms were mainly from the lack of domain knowledge, the misunderstanding of license terms, and the flexibility of defining specific terms. For instance, \textit{Use Trademark} is interpreted as the right to use trademarks, contributor’s name, and logo. However, some participants focus only on the literal word "trademark". For \textit{Compensate for Damages}, it is easy to confuse it with
\textit{Hold Liable}, where the responsible subject is the user, while \textit{Hold
Liable} is the licensor. Moreover, some term (\textit{Modify} and \textit{Commercial Use}) attitudes are not mentioned in their texts but are mentioned in their names, which
is the main reason they are mislabeled.

The participants did not make the labeling errors that occur on other platforms due to confusion of rights and obligations. For the terms \textit{Hold Liable}, \textit{Place Warranty}, \textit{Include Copyright}, and \textit{Include License}, they achieve an average accuracy of 97\%, whereas on other platforms they are heavily mislabeled due to aggregation of terms in~\Cref{subsec:term_identification}, which confirms the quality of our license labeling. Moreover, we have reported some of the mislabeled licenses to relevant platforms, and we received their confirmation that they need more time for further validation. We will update the progress on our website~\cite{ourwebsite}.

\begin{figure}
\footnotesize
\centering
\begin{minipage}{0.4\linewidth}
\captionof{table}{Conflicted attitude by Xu et al.~\cite{xu2021lidetector}}
\scalebox{0.8}{
\begin{tabular}{ll}
\toprule
\textbf{Mode}  & \textbf{Conflicted Attitudes} \\ 
\midrule
\multirow{4}*{\textbf{PL-CL}} & CAN $\xrightarrow{}$ CANNOT  \\
 & CAN $\xrightarrow{}$ MUST    \\ 
 & CANNOT $\xrightarrow{}$ MUST \\
 & MUST $\xrightarrow{}$ CANNOT \\ 
\midrule
\multirow{2}*{\textbf{CL-CL}} & CANNOT $\leftrightarrow{}$ MUST \\
& MUST $\leftrightarrow{}$ CANNOT \\ 
\bottomrule
\end{tabular}}
\label{tab:pre_conflict}
\end{minipage}
\hfill
\begin{minipage}{0.55\linewidth}
\captionof{table}{Conflicted attitude by our rules}
\scalebox{0.8}{
\begin{tabular}{ll}
\toprule
\textbf{Mode}  & \textbf{Conflicted Attitudes} \\ 
\midrule
\multirow{5}*{\textbf{PL-CL}} & CAN $\xrightarrow{}$ CANNOT (for right)\\
 & Not Mentioned $\xrightarrow{}$ MUST\\ & (for obligation)\\
 & CANNOT $\xrightarrow{}$ CAN\\ &  (for copyleft licenses) \\
  \midrule
 \textbf{CL-CL} & N.A. \\
\bottomrule
\end{tabular}}
\label{tab:our_conflict}
\end{minipage}
\begin{enumerate}[leftmargin=*]
    \item $\xrightarrow{}$ indicates the order of projects and components, the source indicates the term attitude of project license, and the sink indicates the term attitude of component license. $\leftrightarrow{}$ indicates the equal relation. 
\end{enumerate}
\end{figure}

\section{License Conflict Analysis}
\label{sec:licenseconflict}

In this section, we further analyze the potential conflict among these SPDX licenses. For reusing third-party components, Xu et al.~\cite{xu2021lidetector} have defined 
license compatibility in two scenarios. \ding{172} A project license (PL) is be ``one-way compatible" with a component license (CL) if anyone who conforms to the PL will not violate CL (\textbf{PL-CL}). \ding{173} Two CLs are compatible if it is possible to develop a new license that anyone who conforms to it will not violate two CLs (\textbf{CL-CL}). Based on these, they have concluded 6 types of license conflict pairs of attitudes as presented in~\Cref{tab:pre_conflict}.

However, their classification neglects the different types of attitudes related to rights and obligations. Specifically, \ding{172} only MUST is assigned to the obligation terms since users are only required to fulfill the obligations; 
\ding{173} CAN and CANNOT are only related to the right terms since rights would never be compulsory. In this case, the conflict between MUST and CANNOT cannot exist (we have confirmed with Xu et al.~\cite{xu2021lidetector} that such conflicts they detected are due to the mislabeling of license terms from third-party platforms). For the conflict CAN $\xrightarrow{}$ MUST, Xu et al.~\cite{xu2021lidetector} simply label unmentioned obligations as CAN, we revised it as NOT MENTIONED $\xrightarrow{}$ MUST for clarity. Moreover, the pair of attitudes CANNOT $\xrightarrow{}$ CAN is treated as compatible in most general cases by Xu et al.~\cite{xu2021lidetector}. However, as vital portions of software licenses, copyleft licenses
are naturally contagious since they require all granted rights preserved in derivative works, thus CANNOT $\xrightarrow{}$ CAN is incompatible when copyleft licenses are assigned to reused components. We also conclude conflicts of copyleft licenses by referring to existing collections~\cite{copyleftlist}.

Based on these, we defined three types of license conflict rules in~\Cref{tab:our_conflict}. \textbf{C1}: For right terms, if a project permits rights that are not granted by component licenses, there could be a conflict.
\textbf{C2}: For obligation terms, if a project does not require obligations required by component licenses, there could be a conflict. \textbf{C3}: If the project licenses fail to preserve all grant rights in components that use the copyleft license, there is a conflict. Note that \textbf{C1} and \textbf{C2} only indicate potential threats (i.e., no legal issues when such cases occur in user projects unless users really violate any terms), while \textbf{C3} is already a real violation of copyleft licenses since the contagion is required by copyleft licenses. Moreover, we ignore the potential conflicts brought by the not-mentioned rights in this paper, since there is still controversy according to the feedback from the OpenChain~\cite{openchain} community (some people think users can do anything that is not forbidden in license, while some think users can only do what is granted in license), but we prefer to preserve this as a future threat. Therefore, 28,918 (\textbf{C1}), 140,870 (\textbf{C2}), and 14,593 (\textbf{C3}) pairs of conflicted licenses are retrieved from the 453 SPDX licenses, respectively.

\noindent \textbf{Validation.} Since C1 and C2 are defined based on existing rules, therefore they are validated as long as the terms and attitude labeling are validated, we further validate the accuracy of \textbf{C3}  by examining our results with publicly known software license conflicts listed by GNU ~\cite{GNU}, which has highlighted licenses that are not incompatible with GNUGPL (i.e., generally it refers to \textit{GPL-3.0-only}). In total, we manually retrieve 40 and 10 SPDX licenses that are explicitly mentioned to conflict with GNUGPL and GFDL, and find that 46 licenses all conflict with \textit{GPL-3.0-only}, and they are all covered by \textbf{C3}. As for the other 4, we find that they are all conflicts between strong copyleft and weak copyleft (i.e., whether derivative works are required to be granted the same license).

To facilitate further studies on the basis of our well-constructed license dataset, all labeled licenses with their terms, as well as the three types of conflicts (\textbf{C1} \textasciitilde \textbf{C3}) are available on our website~\cite{ourwebsite}.

\section{License Conflict Study} \label{subsec:conflictstudy}
We investigate the similarities and differences of licenses to interpret the major concerns of designing so many different licenses by answering the first research question. Specifically, based on 453 SPDX licenses, we first investigate the distribution of rights and obligations that are commonly defined. Next, we investigate the major differences by clustering them via community detection and frequent itemset mining. Last, we further investigate the features of licenses that bring conflicts.
    
\noindent $\bullet$ \textbf{RQ1: License Interpretation.} What are the major terms that consist of SPDX licenses? What are the major differences that bring the diversity of SPDX licenses? What are the major concerns that are easy to cause license conflicts?

\begin{figure*}
\center
\includegraphics[width=0.85\textwidth]{"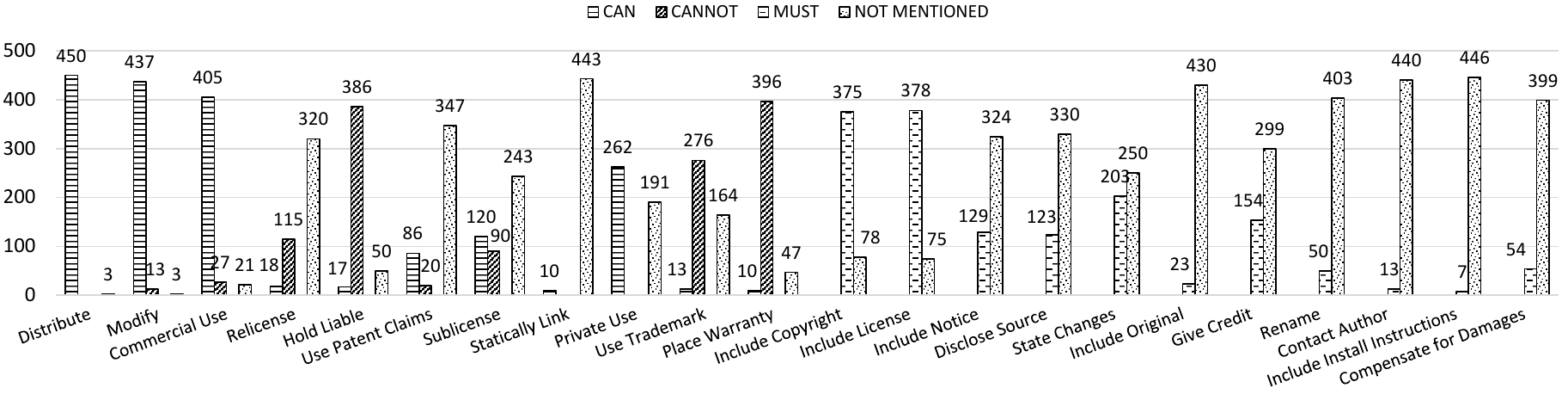"}
\caption{The distribution of rights and obligations in SPDX licenses.}
\label{fig:majorpatterns}
\end{figure*}

\begin{figure}
\centering
\begin{minipage}{0.46\linewidth}
\center
\includegraphics[width=\textwidth]{"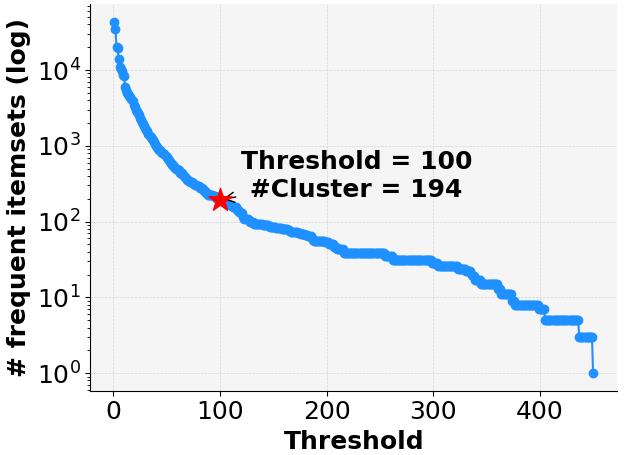"}
\captionof{figure}{The frequent patterns with different thresholds in license terms.}
\label{fig:frequent}
\end{minipage}
\hfill
\begin{minipage}{0.46\linewidth}
\center
\includegraphics[width=\textwidth]{"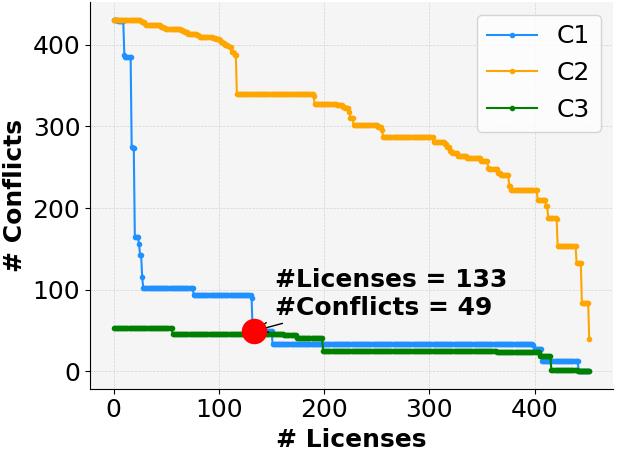"}
\captionof{figure}{The distribution of C1\textasciitilde C3 conflicts for SPDX licenses.}
\label{fig:conflict}
\end{minipage}
\end{figure}

\subsection{{Major Terms and Their Attitudes}} \label{subsubsec:majorterms}
First, we analyze the existence of granted rights and required obligations in each SPDX license. As presented in~\Cref{fig:majorpatterns}, the license terms are not always included in all the 453 SPDX licenses. Specifically, most license requirements \textit{Include License} (378) and \textit{Include Copyright} (375) when licensees are granted the rights of \textit{Distribute} (450) and \textit{Modify} (437) software, and meanwhile, they refuse to \textit{Place Warranty} (396) and \textit{Hold Liable} (386) for the distributed software. Furthermore, most SPDX licenses also permit \textit{Commercial Use} (405) for better distribution. For better presentation, we name the commonly mentioned terms (i.e., \textit{Distribute, Modify, Commercial Use, Hold Liable, Place Warranty, Include Copyright}, and \textit{Include License}) as \textbf{Common Term Set}.

Apart from these common terms that are usually defined with the same attitude, we also identify terms that are seldom mentioned in mainstream licenses. For instance, as a special scenario, \textit{Statically Link} is gradually raised when clarifying the rights of usage, while none of these licenses explicitly grants this right, and it is only indirectly included in the 10 licenses that grant all rights. \textit{Relicense} (18 CAN and 115 CANNOT) is also seldom included and permitted in licenses. Such a low permission rate makes it difficult to resolve license incompatibility issues after long-term collaborations. For instance, LLVM has proposed to change the license from NCSA license to Apache 2.0 since 2015, but this process has not been finished yet due to \textit{Relicense} is not granted, it is necessary to seek approval from all contributors~\cite{llvmrelicense}. 

Furthermore, \textit{Include Original}, \textit{Rename}, \textit{Contact Author} and \textit{Include Install Instructions} are also unpopular in SPDX licenses. {\textit{Include Original}, \textit{Rename}, and \textit{Include Install Instructions} increase the additional efforts from licensees, and \textit{Contact Author} can be annoying because the licensees have to contact the authors for additional permissions. These terms hinder the distribution and reuse of software, which is why we believe they are unpopular in modern software licenses.

\subsection{{License Differences}} \label{subsubsec:licensediversity}
Based on these most general terms in SPDX licenses, we also inspect the common patterns across terms. Specifically, in order to identify the license differences, we summarize the high co-existence of license terms by FP-mining~\cite{frequentmining} to interpret their common features. Considering the rapid growth of mining patterns and duplication, we set the threshold of FP-Growth algorithm~\cite{fpgrowth} to 100, as shown in~\Cref{fig:frequent}. We identify 196 frequent patterns of terms that exist in over 100 SPDX licenses. After excluding similar patterns (over 90\% common licenses), we identify 5 groups of the most common patterns. For better clarification, we skip these features that coexist with the \textbf{Common Term Set} as they usually have a high co-existence rate with almost all other license terms, and we interpret the major differences of each group as follows. 

\noindent \textbf{(1) Private Use \& State Changes (118 licenses).} Both of \textit{Private Use} (262) and \textit{State Changes} (203) have relatively high mentions. Granting \textit{Commercial Use} naturally also grants \textit{Private Use}, this inclusive relationship is reflected in our data. For \textit{State Changes}, 188 licenses that grant \textit{Commercial Use} also require \textit{State Changes} as a precondition, and interestingly only 117 of these licenses also explicitly grant \textit{Private Use}. Therefore, we believe that a more precise and formal model of license terms should be further concluded.

\noindent \textbf{(2) Sublicense (120 licenses).} As one of the most concerned rights, \textit{Sublicense} allows the licensees to grant permitted rights to third parties. We notice \textit{Sublicense} is always granted with \textit{Distribute}, \textit{Modify}, and \textit{Commercial Use} (in 119 licenses), while the obligations differ in this group, only 84 licenses contain both \textit{Sublicense} and the \textbf{Common Term Set}.
    
\noindent \textbf{(3) Disclose Source (123 licenses).} This obligation requires the source code must be available when the software is distributed, indicating a premise condition when granting the right of \textit{Distribute}, and all licenses with \textit{Disclose Source} are also granted with \textit{Distribute}, while only 27.3\% of licenses with \textit{Distribute} has mentioned such premise condition.
    
\noindent \textbf{(4) Include Notice (129 licenses).} Some licenses also highlighted the obligation to include relevant notices when distributing the work. We notice that it has a high co-existence with \textit{State Changes} (in 90 licenses), which requires stating the changes made to the work, as a typical notice. 

\noindent \textbf{(5) Give Credit (154 licenses).} This is another prerequisite for distributing software to explicitly give credit to the authors. The obligation is usually declared in Creative Commons licenses \cite{Creativecommons} and document licenses (i.e., GFDL licenses~\cite{GFDL}), as an exchange for granting rights, and after version 2.0, all CC licenses are required to \textit{Give Credit}.

Overall, although 22 license terms are commonly defined to build software licenses, apart from the \textbf{Common Term Set}, only limited terms, in particular the obligation terms specifying different purposes, derive the major diversity of licenses. 

\subsection{{Major Conflicts}}
In this section, we further interpret the diversity of SPDX licenses from the perspective of major conflicts. For \textbf{C1}~\textasciitilde \textbf{C3}, we separately examine the incompatible license pairs identified in~\Cref{sec:licenseconflict}, and find out the licenses that conflict with most other licenses. Moreover, we also identify the major terms that introduce these license conflicts.

\Cref{fig:conflict} presents the distribution of licenses that have \textbf{C1} \textasciitilde \textbf{C3} conflicts with others. 
\textbf{C2} is the primary type of conflict among SPDX licenses, over 400 licenses have obligation conflicts with at least 200 other licenses. The conflicts caused by \textbf{C1} are much fewer except for the top 28 licenses that have different right permissions with over 100 other licenses. Moreover, 321 licenses only have rights conflicts with less than 50 other licenses. Note that both \textbf{C1} and \textbf{C2} do not indicate the violation of granted licenses when projects and their components adopt conflicting licenses, but such an incredible amount of potential conflicts makes it necessary to examine the licenses of software dependencies, especially transitive dependencies, and to strictly adhere to the rights granted and obligations required by each license in dependencies.

Moreover, we investigate the major conflicted terms in all \textbf{C1} \textasciitilde \textbf{C3} conflicts. According to our observation, as the most common concerned terms, \textit{Commercial Use} and \textit{Sublicense} contribute the most C1 conflicts (37\% for both). \textit{Hold Liable} and \textit{Place Warranty} also contribute 23\% and 13\% C1 conflicts, respectively, since many licenses explicitly discard such rights to licensees to avoid further responsibilities. For C2 conflicts, \textit{State Changes} (36\%), \textit{Give Credit} (32\%), \textit{Include Notice} (30\%) and \textit{Disclose Source} (29\%) are the top obligation terms that bring the most conflicts, which are consistent with the mostly concerned obligation terms that are related to premise conditions of \textit{Distribute}, as we unveiled in~\Cref{subsubsec:licensediversity}. For C3 conflicts, we surprisingly find that \textit{Use Patent Claims} contributes the most C3 conflicts (58\%). We further go through these copyleft licenses, and find that most copyleft and public licenses (except \textit{GPL-family} licenses) have granted the right of \textit{Use Patent Claims}, such as EPL~\cite{epl}, EUPL~\cite{eupl}, IPL~\cite{ipl}, LPL~\cite{lpl}, MPL~\cite{mpl}, OSL~\cite{osl}, RPL~\cite{rpl}, etc. We speculate that copyleft licenses generally require derivative work of open-source software to be distributed under the same license. By granting \textit{Use Patent Claims}, the potential legal disputes could be easily avoided.

\begin{tcolorbox}[size=title,opacityfill=0.2,breakable,boxsep=1mm]
\textbf{Answers to RQ1:} There exists clear centrality on license terms and their corresponding attitudes in the mainstream licenses. \ding{172} \textit{Distribute, Modify, Commercial Use}, \textit{Place Warranty, Hold Liable}, \textit{Include License}, and \textit{Include Copyright} are the mostly common terms mentioned in software licenses, indicating high consistency on basic requirements of reusing software artifacts. \ding{173} Apart from \textit{Sublicense}, the major differences of mainstream licenses mostly lay in the obligations contained in the preconditions for \textit{Distribute}, such as \textit{State Changes, Disclose Source, Include Notice,} and \textit{Give Credit}. \ding{174} These major differences also contribute the most term-based license conflicts (i.e., \textbf{C1} and \textbf{C2}), while the majority of copyleft conflicts (\textbf{C3}) result from \textit{Use Patent Claims}, which is usually not included in most copyright licenses.
\end{tcolorbox}

\section{NPM License Study}
\label{sec:Empirical}

In this section, we revisit the NPM ecosystem, one of the largest ecosystems, that has been studied and claimed to have less conflict risk by previous work~\cite{ShiQiu2021empirical}, to peek into the license adoption and potential conflicts with the basis of fine-grained license labeling. Specifically, we conduct the study by answering the following research questions:

\noindent $\bullet$ \textbf{RQ2: License Usage and Conflicts in the NPM Ecosystem.} What are the minority licenses except for MIT in the NPM ecosystem? Do maintainers change their licenses throughout the maintenance of NPM packages? Does adopting the common \textit{MIT} license really protect maintainers from potential license violations?

\noindent \textbf{Data Preparation.} We collect all the packages published on the NPM registry \cite{NPM} until June 2023. In total, 3.37 million libraries, including 35.91 million versions, are retrieved. Based on the metadata, we collect the license assigned to each package. Considering that not all licenses are defined in a standardized way in the metadata, we manually inspect and try to unify them to SPDX identifiers. Finally, we successfully collect the licenses of 3.25 million libraries and 32.93 million versions. The unparsable licenses are mainly those not in SPDX licenses or stated in ambiguous text (e.g., file paths, URLs, or hash codes). Moreover, we also collect the dependencies of each library version for license conflict and evolution study.

\subsection{\textbf{The Minority Licenses}} 
We first revisit the distribution of license usage in the NPM ecosystem. As revealed by existing works~\cite{pfeiffer2022license, ShiQiu2021empirical, makari2022prevalence}, \textit{MIT} is the most popular license in the NPM ecosystem, half of libraries use \textit{MIT} license. These works generally focus only on the most commonly used licenses, while the minority licenses may also raise compatibility threats. Therefore, we focus on the less popular licenses adopted by NPM libraries. Specifically, we collect the licenses that are used by over 100 libraries. Apart from \textit{MIT} (1.20m, 37\%), \textit{ISC} license (1.33m, 41\%) is also widely used by NPM packages, probably because \textit{ISC} is the default license assigned by the NPM ecosystem when releasing packages. We also find that 17\% of libraries have no license (0.56m), which is surprising since users manually deleted the default \textit{ISC} licenses assigned by NPM. Besides them, we identify 35 other minority licenses that are used by the remaining 156K libraries (4.8\%).

Moreover, similar to \textit{MIT} and \textit{ISC} licenses, another 12 permissive licenses take the majority of the rest licenses, which accounts for 132K libraries. Among them, \textit{Apache-2.0} is the third most used license in the NPM ecosystem. We also find 13 copyleft licenses in the minority licenses. Specifically, 14.8K libraries use the \textit{GPL-family} licenses, and 5.4K libraries choose other copyleft public licenses. Note that these copyleft licenses may bring severe license incompatibility issues (\textbf{C3} conflicts as discussed in~\Cref{sec:licenseconflict}). 

\begin{figure}
\center
\includegraphics[width=0.35\textwidth]{"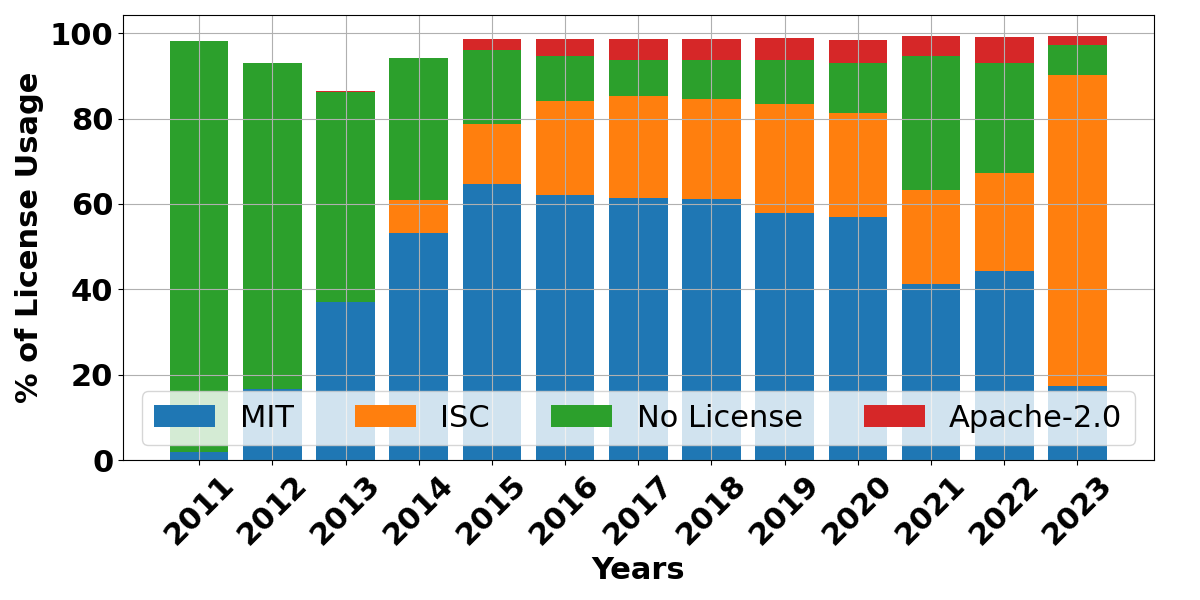"}
\caption{The evolution of the major license usage in the NPM.}
\label{fig:usage}
\end{figure}

\subsection{License Changes} 
We first investigate the evolution of license usage. Specifically, we retrieve the license for the latest version of each library released each year. \Cref{fig:usage} presents the evolution of the usage of major licenses in the NPM ecosystem. Before 2014, the majority of NPM packages were released without a license, and the MIT license was only used to a small extent. After \textit{ISC license} was set as the default license when publishing packages via NPM, it gradually became one of the most popular licenses~\cite{makari2022prevalence}. Even though we note the rapid increase in packages with no license after 2021, it has no effect on the fact that ISC license suddenly accounts for 70\% of released NPM packages in 2023. We further inspect a sample of these recently released packages with no license or ISC license and find that most of them are either sudden releases after years of inactivity or new packages with only one version, which matches the SEO spam attack~\cite{spam} against the NPM ecosystem in March, because most of the auto-released packages neglect to select a proper license.

Moreover, we also inspect the license changes made by maintainers. Generally, NPM maintainers prefer to use a consistent license across their releases. In total, 67,953 libraries have changed their license when publishing new versions (79,446 times in total). After excluding changes to license statements rather than license content, (e.g., changing from \textit{SEE LICENSE IN LICENSE.TXT} to a specific license), and abnormal license names (e.g., hash codes or URLs), we obtain 59,973 pairs of license changes. We find that 74.7\% of license changes (44,811) are from a permissive license to another permissive license, of which 29,918 (49.9\%) to MIT and the rest to \textit{ISC} (7,543), \textit{Apache-2.0} (5,072) and \textit{BSD} (2,278). Based on the data we can see that maintainers have a clear preference for \textit{MIT} license because it explicitly grants \textit{Sublicense} compared to other permissive licenses, which better serve user demands. Moreover, we also note that 3,251 and 2,243 license changes are between permissive licenses and copyleft licenses, and the majority of these changes (3,749) are from packages released by organizations rather than individuals, which also confirms the infectious requirement of the Copyleft licenses. In this case, it is still vital to monitor the license conflicts, especially for packages that are fundamental libraries.


\begin{table}[]
\caption{Top-10 conflicted license pairs in direct dependencies for NPM}
\scalebox{0.45}{
\begin{tabular}{llc|llc|llc}
\toprule
\multicolumn{3}{c|}{\textbf{C1}}             & \multicolumn{3}{c|}{\textbf{C2}}              & \multicolumn{3}{c}{\textbf{C3}}                \\
\midrule
Parent & Dependency & \# & Parent & Dependency & \# & Parent & Dependency & \# \\
\midrule
MIT & CC0-1.0             & 109,659 & MIT & Apache-2.0          & 3,014,374 & MIT & MPL-2.0                & 82,200  \\
MIT&GPL-3.0-or-later    & 24,979  & ISC & Apache-2.0          & 1,166,969 & MIT & GPL-3.0-or-later       & 24,979  \\
MIT&GPL-2.0-or-later    & 22,268  & Unlicense&MIT           & 97,568   & ISC & MPL-2.0                & 22,262  \\
Apache-2.0 & CC0-1.0      & 20,283  & MPL-2.0 & Apache-2.0      & 95,712   & ISC & GPL-3.0-or-later       & 9,477   \\
Apache-2.0 & CC-BY-ND-4.0 & 14,876  & MIT & MPL-2.0             & 82,200   & LGPL-3.0-only & GPL-3.0-only & 1,982   \\
MIT & CC-BY-4.0           & 14,847  & BSD-3-Clause & Apache-2.0 & 80,831   & AAL & MPL-2.0                & 1,606   \\
CECILL-B & MIT            & 6,347   & WTFPL & MIT               & 59,052   & LGPL-3.0-only & MPL-2.0      & 1,360   \\
MPL-2.0 & CC0-1.0         & 6,233   & MIT & WTFPL               & 53,477   & BSD-3-Clause & MPL-2.0       & 1,278   \\
Apache-2.0 & GPL-3.0-only & 5,178   & CC0-1.0&MIT             & 49,628   & MIT & EPL-1.0                & 1,223   \\
MIT & CC-BY-3.0           & 5,018   & MIT & Artistic-2.0        & 32,539   & ISC & GPL-3.0-only           & 1,193   \\
\midrule
Total         &            & 229,688 & Total         &            & 4,732,350 & Total              &          & 147,560 \\
\bottomrule
\end{tabular}}
\label{tbl:topconflict}
\end{table}

\subsection{License Conflicts} 
We further analyze the potential license conflicts brought by dependencies. We use the dependency parser from Liu et al.~\cite{liu2022demystifying} to parse the dependency ranges to specific versions. Over 228 million direct dependencies have been identified and collected, of which 5.33 million (2.3\% of direct dependencies) are identified as having potential license conflicts. Specifically, 5.18m of these license conflicts are obligation conflicts (\textit{C2}), right (\textit{C1}) and copyleft (\textit{C3}) conflicts account for 261K and 159K respectively. Note that some dependencies may have more than one type of conflict.

\Cref{tbl:topconflict} presents the top-10 license pairs with the most \textbf{C1} \textasciitilde \textbf{C3} conflicts in dependencies. For \textbf{C1} conflicts, the \textit{MIT} license contributes the most right conflicts. In particular, conflicts triggered when packages are released under the \textit{MIT} license and their dependencies are with a more restrictive license (177K in total, 67.8\%) account for 5 of the top 10 conflict pairs. For instance, 109K C1 conflicts are between \textit{MIT} and \textit{CC0-1.0}, and the conflict is because \textit{MIT} permits \textit{Sublicense} while \textit{CC0-1.0} forbids it. For \textbf{C2} conflicts, \textit{Apache-2.0} contributes the most conflicts. Specifically, 4.42m C2 conflicts (85.3\%) are introduced when dependencies are released with \textit{Apache-2.0}, even if these libraries are released under the \textit{MIT} or \textit{ISC}. These conflicts are mainly because \textit{Apache-2.0} requires \textit{Include Notice} and \textit{State Changes}. So although \textit{Apache-2.0} is a permissive license, users still need to be aware of these additional obligations. For \textbf{C3} conflicts, since copyleft licenses are naturally infectious, conflicts occur when the rights granted are not included in the parent license. The most common conflict term \textit{Use Patent Claims} in C3, causes almost all C3 conflicts (157K, 98.5\%). 
Therefore, although permissive licenses are dominant in the NPM ecosystem, we still call for more attention to the more restrictive licenses and copyright licenses hidden in the dependencies.

\begin{tcolorbox}[size=title,opacityfill=0.2,breakable,boxsep=1mm]
\textbf{Answers to RQ2:} 
The extensive usage of permissive licenses in the NPM ecosystem does not indicate less risk. \ding{172} Apart from the most common permissive licenses (e.g., \textit{MIT} and \textit{ISC}), there are a number of less popular permissive licenses (in 132 libraries) and copyleft licenses (e.g., 15K \textit{GPL-family} licenses) that are also widely used. \ding{173} Maintainers seldom change their licenses unless special needs are not met. \ding{174} Although most license conflicts are raised by minority licenses, they are still related to common licenses. For instance, 67.8\% of right conflicts are related to \textit{Sublicense} granted in the \textit{MIT}, and 85.3\% of obligation conflicts result from obligations (\textit{Include Notice} and \textit{State Changes}) not required in the MIT. 
\end{tcolorbox}

\section{Discussion}
\label{discussion}
\subsection{Lessons learned}
\textbf{For Developers}. Since developers as both providers and consumers of software packages, we conclude with some tips for them: 1) Developers should not offer software packages without a license as there are significant legal risks. 2) Developers should check license information and judge compatibility when integrating third-party packages. 3) Community developers should routinely examine licensing conflicts on dependencies at scale to eliminate or minimize the impact on indirect dependencies. Our datasets help developers quickly understand license content and conflicts, guiding developers to make initial decisions about license selection and compatibility before consulting with lawyers. In addition, developers can develop automated tools for license identification, term extraction, and conflict judgment based on our datasets. 


\textbf{For Researchers}. Licenses will continue to evolve with the development of open-source software, so research on licenses will also continue. We conclude with some tips: 1) Researchers can refer to some data from existing platforms, but should not use them directly because their quality needs to be verified. 2) It is not enough for researchers to focus on term-based license conflict, because there are inherently incompatible license pairs that cannot be judged by terms, such as Copyleft licenses, researchers should continue to refine license conflict models. 3) Rights or obligations not mentioned in the license cannot be directly ignored just because they are controversial but should be preserved as a future threat. Overall, some of the stereotypes about licenses still need to be broken by extensive empirical studies by researchers.


\subsection{Limitations and threats to validity}
\label{sec:Limitation}
1) We identify 22 license terms based on a comparative analysis of data from three mainstream platforms and the textual content of 453 licenses. However, there could be specific requirements beyond these 22 terms.
2) Licenses in NPM are presented in various ways, some libraries are ignored in our study. However, we have tried our best to manually check for abnormal license names, and only less than 3\% libraries are neglected.
3) We ignore the potential conflicts brought by the not-mentioned rights since there is still controversy according to the feedback from the OpenChain~\cite{openchain} community, but we have kept them as a future threat. 
4) We cannot guarantee the seriousness of all questionnaire participants. However, only one result is discarded due to not following the requirements, and the rest results are highly consistent. 

\section{Related Work}
\textbf{License Interpretation.} 
Alspaugh et al.~\cite{Intellectualproperty, Softwarelicensesin} extracted tuple information from license text as terms to model 10 licenses. 
Ballhausen et al.~\cite{ballhausen} explained free and open-source software licenses in license terms. 
Gordon et al.~\cite{Qualipso} modeled eight popular licenses by ontology. 
Kapitsaki et al.~\cite{kapitsaki2017identifying, Kapitsaki2019modeling} retrieved the properties defined in 24 licenses via topic models. Cui et al.~\cite{cui2023study} defined 12 terms by extracting information from a large number of custom licenses via NLP. 

Some existing works focus on the actual use of licenses. Almedia et al.~\cite{almeida} investigated whether developers really understand open-source licenses. Coelho et al.~\cite{survey} surveyed GitHub contributors and confirmed the importance of proper usage of licenses. Chris et al.~\cite{jensen2011license} investigated the license update and migration processes in open-source projects. Yuki et al.~\cite{evolutional} surveyed the evolution of several mainstream licenses. Maria et al.~\cite{Maria} proved that practitioners need clarifications about licensing specific software when other licenses are used.

\textbf{License Compatibility Analysis.}
Wheeler et al.~\cite{Thefree-libre} manually constructed a directed graph to represent the compatibility between commonly used licenses. Gordon et al.~\cite{gordon2021analyzing} proposed a legal prototype to analyze the license compatibility of Carneades, while it did not detect real-world compatibility. 
Kapitsaki et al.~\cite{kapitsaki2015open, kapitsaki2017automating} detected license incompatibility based on a manually built compatibility graph, but it only covered a small set of licenses. Xu et al.~\cite{xu2021lidetector} proposed a NER-based tool LiDetector to retrieve terms and detect incompatibility for customized licenses. Moreover, they also proposed LiResolver~\cite{xu2023liresolver} to resolve license incompatibility based on LiDetector. However, LiDetector is trained on labeled data provided by TLDDRLegal~\cite{tldr}, which largely harms the accuracy, as discussed in~\Cref{subsec:licenselabeling}.

Some works also investigate license incompatibility in software ecosystems.
Qiu et al. \cite{ShiQiu2021empirical} studied NPM software packages based on a small compatibility network built and found that only 0.644\% of packages had dependency-related license violations. However, their conclusion was also largely compromised by the small compatibility network.
Joao et al.~\cite{moraes2021one} investigated the multi-licenses in the JavaScript ecosystem. 
Makari et al.~\cite{makari2022prevalence} studied the evolution, popularity, and compliance with dependency licenses in the NPM and RubyGems. 
Pfeiffer et al. \cite{pfeiffer2022license} studied license usage and incompatibility between components from seven package registries. 


Many works suffer from low-quality data on license, in contrast, we construct a high-quality dataset of 453 SPDX licenses, covering 97\% of licenses used in NPM. 

\section{Conclusion}

In order to figure out the disorder of interpretation for software licenses in the open source community, we conduct the first large-scale analysis based on SPDX licenses to clarify the granted permissions and required obligations by differential analysis, including proposing a standardized license term set, labeling 453 SPDX licenses. Moreover, we reveal three types of license conflicts (summarizing for the first time the conflicts introduced by the infectious copyleft licenses) and investigate the license terms and corresponding attitudes that cause the main differences and conflicts between them. Building on this foundation, we further carry out an ecosystem-wide study to revisit the license usage and conflicts in the NPM ecosystem. Our study confirms the predominance of permissive licenses in NPM. In addition, our study reveals that while most maintainers prefer keeping licenses unchanged, the minority libraries that use restrictive licenses and copyleft licenses introduce excessive potential conflicts.

\section*{Acknowledgements}
This work was supported by the National Key Research and Development Program (2023YFB3106400, 2023QY1202), the National Natural Science Foundation of China (U2336203, U1836210), the Key Research and Development Science and Technology of Hainan Province (GHYF2022010), and Tao Liu was also funded by China Scholarship Council. This work was also supported by the National Research Foundation, Singapore, and DSO National Laboratories under the AI Singapore Programme (AISG Award No: AISG2-GC-2023-008), and the Cyber Security Agency under its National Cybersecurity R\&D Programme (NCRP25-P04-TAICeN) and the NRF Investigatorship NRF-NRFI06-2020-0001. Any opinions, findings and conclusions or recommendations expressed in this material are those of the author(s) and do not reflect the views of National Research Foundation, Singapore and Cyber Security Agency of Singapore.

\bibliographystyle{ieeetr}
\bibliography{ref}
\end{document}